\documentclass[longauth]{aa} 
\usepackage{graphicx}
\usepackage{txfonts}
\usepackage{natbib}
\bibpunct{(}{)}{;}{a}{}{,} 

\def\ms{\hbox{m\,s$^{-1}$}}         
\def\m2s2{\hbox{m$^{2}$\,s$^{-2}$}} 
\def\kms{\hbox{km\,s$^{-1}$}}       
\def\gcm3{\hbox{g\,cm$^{-3}$}}      
\def\Mjup{\hbox{$\mathrm{M}_\mathrm{J}$}}
\def\Rjup{\hbox{$\mathrm{R}_\mathrm{J}$}}
\def\degr{\hbox{$^\circ$}}

\begin{document}
   \title{Transiting exoplanets from the CoRoT space mission}

   \subtitle{XI. CoRoT-8b: a hot and dense sub-Saturn around a K1 dwarf\thanks{Observations made with SOPHIE spectrograph at Observatoire de
   Haute Provence, France (PNP.07B.MOUT), and the HARPS spectrograph at ESO La Silla Observatory (081.C-0388 and 083.C-0186).
   The CoRoT space mission, launched on December 27, 2006, has been developed and is operated by the CNES with the contribution
   of Austria, Belgium, Brasil, ESA, Germany, and Spain.}}

   \author{P.~Bord\'e\inst{\ref{IAS}}                     \and F.~Bouchy\inst{\ref{IAP},\ref{OHP}}    \and M.~Deleuil\inst{\ref{LAM}}
           \and J.~Cabrera\inst{\ref{DLR},\ref{LUTH}}     \and L.~Jorda\inst{\ref{LAM}}               \and C.~Lovis\inst{\ref{Geneva}}
           \and S.~Csizmadia\inst{\ref{DLR}}              \and S.~Aigrain\inst{\ref{Oxford}}          \and J.~M.~Almenara\inst{\ref{IAC},\ref{Tenerife}}
           \and R.~Alonso\inst{\ref{Geneva}}              \and M.~Auvergne\inst{\ref{LESIA}}          \and A.~Baglin\inst{\ref{LESIA}}
           \and P.~Barge\inst{\ref{LAM}}                  \and W.~Benz\inst{\ref{Bern}}               \and A.~S.~Bonomo\inst{\ref{LAM}}
           \and H.~Bruntt\inst{\ref{LESIA}}               \and L.~Carone\inst{\ref{Koln}}             \and S.~Carpano\inst{\ref{ESTEC}}
           \and H.~Deeg\inst{\ref{IAC},\ref{Tenerife}}    \and R.~Dvorak\inst{\ref{Vienna}}           \and A.~Erikson\inst{\ref{DLR}}
           \and S. Ferraz-Mello\inst{\ref{SaoPaulo}}      \and M.~Fridlund\inst{\ref{ESTEC}}          \and D.~Gandolfi\inst{\ref{ESTEC},\ref{TLS}}
           \and J.-C.~Gazzano\inst{\ref{LAM}}             \and M.~Gillon\inst{\ref{IAG}}              \and E.~Guenther\inst{\ref{TLS}}
           \and T.~Guillot\inst{\ref{OCA}}
           \and P.~Guterman\inst{\ref{LAM}}               \and A.~Hatzes\inst{\ref{TLS}}              \and M.~Havel\inst{\ref{OCA}}
           \and G.~H\'ebrard\inst{\ref{IAP}}              \and H.~Lammer\inst{\ref{Graz}}             \and A.~L\'eger\inst{\ref{IAS}}
           \and M.~Mayor\inst{\ref{Geneva}}               \and T.~Mazeh\inst{\ref{Wise}}              \and C.~Moutou\inst{\ref{LAM}}
           \and M.~P\"atzold\inst{\ref{Koln}}             \and F.~Pepe\inst{\ref{Geneva}}             \and M.~Ollivier\inst{\ref{IAS}}
           \and D.~Queloz\inst{\ref{Geneva}}              \and H.~Rauer\inst{\ref{DLR},\ref{Berlin}}  \and D.~Rouan\inst{\ref{LESIA}}
           \and B.~Samuel\inst{\ref{IAS}}                 \and A.~Santerne\inst{\ref{LAM}}            \and J.~Schneider\inst{\ref{LUTH}}
           \and B.~Tingley\inst{\ref{IAC},\ref{Tenerife}} \and S.~Udry\inst{\ref{Geneva}}             \and J.~Weingrill\inst{\ref{Graz}}
           \and G.~Wuchterl\inst{\ref{TLS}}
          }

   \institute{Institut d'astrophysique spatiale, Universit\'e Paris-Sud 11
              \& CNRS (UMR 8617), B\^at. 121, 91405 Orsay, France \label{IAS} \\
              \email{pascal.borde@ias.u-psud.fr}
         \and 
             Institut d'astrophysique de Paris, Universit\'e Paris 6
             \& CNRS (UMR 7095), 98 bd Arago, 75014 Paris, France \label{IAP}
         \and 
             Observatoire de Haute-Provence, CNRS \& OAMP,
             04870 St-Michel l'Observatoire, France \label{OHP}
         \and 
             Laboratoire d'astrophysique de Marseille, Universit\'e de Provence
             \& CNRS (UMR 6110), 38 rue F. Joliot-Curie, 13388 Marseille, France \label{LAM}
         \and 
             Institute of Planetary Research, German Aerospace Center, Rutherfordstrasse 2,
             12489 Berlin, Germany \label{DLR}
         \and
             Laboratoire de l'univers et de ses th\'eories, Observatoire de Paris
             \& CNRS (UMR 8102), 5 place Jules Janssen, 92195 Meudon, France \label{LUTH}
         \and 
             Observatoire de Gen\`eve, Universit\'e de Gen\`eve,
             51 Ch. des Maillettes, 1290 Sauverny, Switzerland \label{Geneva}
         \and
             Department of Physics, Denys Wilkinson Building Keble Road, Oxford, OX1 3RH \label{Oxford}
         \and 
             Instituto de Astrof\'isica de Canarias, E-38205 La Laguna, Tenerife,
             Spain \label{IAC}
         \and
             Departamento de Astrof\'isica, Universidad de La Laguna, E-38200
             La Laguna, Tenerife, Spain \label{Tenerife}
         \and
             Laboratoire d'\'etudes spatiales et d'instrumentation en astrophysique,
             Observatoire de Paris \& CNRS (UMR 8109), 5 place Jules Janssen,
             92195 Meudon, France \label{LESIA}
         \and
             Physikalisches Institut Universit\"at Bern, Sidlerstrasse 5,
             3012 Bern, Switzerland \label{Bern}
         \and 
             Rheinisches Institut f\"ur Umweltforschung an der Universit\"at zu K\"oln,
             Aachener Str. 209, 50931 K\"oln, Germany \label{Koln}
         \and 
             Research and Scientific Support Department, European Space Agency,
             Keplerlaan, NL-2200AG, Noordwijk, The Netherlands \label{ESTEC}
         \and 
             Institute for Astronomy, University of Vienna, T\"urkenschanzstrasse 17,
             1180, Vienna, Austria \label{Vienna}
         \and
             Institute of Astronomy, Geophysics and Atmospheric Sciences,
             University of S\~ao Paulo, Brasil \label{SaoPaulo}
         \and 
             Th\"uringer Landessternwarte, 07778 Tautenburg, Germany \label{TLS}
         \and 
             IAG, Universit\'e de Li\`ege, All\'ee du 6 ao\^ut 17, Li\`ege 1, Belgium \label{IAG}
         \and
             Universit\'e de Nice-Sophia Antipolis, CNRS UMR 6202, Observatoire de la C\^ote d'Azur,
             BP 4229, 06304 Nice Cedex 4, France \label{OCA}
         \and
             Space Research Institute, Austrian Academy of Sciences,
             Schmiedlstr. 6, Graz, Austria \label{Graz}
         \and 
             Wise Observatory, Tel Aviv University, Tel Aviv 69978, Israel \label{Wise}
         \and             
             Center for Astronomy and Astrophysics, TU Berlin, Hardenbergstrasse 36,
             10623 Berlin, Germany \label{Berlin}
             }

   \date{Received ?; accepted ?}

 
  \abstract
   {} 
   {We report the discovery of CoRoT-8b, a dense small Saturn-class exoplanet that orbits a K1
   dwarf in 6.2 days, and we derive its orbital parameters, mass, and radius.}
   {We analyzed two complementary data sets: the photometric transit curve of CoRoT-8b as measured by
   CoRoT and the radial velocity curve of CoRoT-8 as measured by the HARPS spectrometer\thanks{Both data
   sets are available in electronic form at the CDS via anonymous ftp to \texttt{cdsarc.u-strasbg.fr
   (130.79.128.5)} or via \texttt{http://cdsweb.u-strasbg.fr/cgi-bin/qcat?J/A+A/vol/page}.}.}
   {We find that CoRoT-8b is on a circular orbit with a semi-major axis of $0.063 \pm 0.001$~AU. It
   has a radius of $0.57 \pm 0.02$~\Rjup, a mass of $0.22 \pm 0.03$~\Mjup, and therefore a mean density
   of $1.6 \pm 0.1$~\gcm3.}
   {With 67\,\% of the size of Saturn and 72\,\% of its mass, CoRoT-8b has a density comparable to that
   of Neptune (1.76~\gcm3). We estimate its content in heavy elements to be 47--63~$\mathrm{M}_\oplus$,
   and the mass of its hydrogen-helium envelope to be 7--23~$\mathrm{M}_\oplus$. At 0.063~AU,
   the thermal loss of hydrogen of CoRoT-8b should be no more than $\sim 0.1$\,\% over an assumed
   integrated lifetime of 3~Ga.}


   \maketitle


\section{Introduction}

   CoRoT (Convection, Rotation and planetary Transits), a joint project of France, Austria,
   Belgium, Brazil, ESA, Germany, and Spain, is a low-Earth orbit visible photometer designed
   to measure stellar light curves (LCs) for two main astrophysical programs: asteroseismology
   and exoplanet detection \citep{Baglin2006}. Instrument characteristics and inflight
   performance can be found in \cite{Auvergne2009}. CoRoT was launched on December
   27, 2006 and will be operated by the French Centre National d'\'Etudes
   Spatiales (CNES) until March 31, 2013.
   
   The search for exoplanets in CoRoT LCs consists in looking for periodic dips caused
   by exoplanets whose edge-on orbits make them transit in front of their host stars.
   Study of transit signals combined with ground-based radial velocity follow-up
   observations leads to complete orbital solutions, as well as exoplanet masses and
   radii. It is with these systems that we learn the most about exoplanetary internal
   structures, atmospheres, and dynamics \citep[for a review, see][]{Charbonneau2007}.
   Compared to ground-based transit surveys, CoRoT offers uninterrupted photometric
   time series for up to about 150 days with relative errors as low as about 100 ppm per hour.
   This makes it possible to hunt for exoplanets that are only twice as big as the Earth in
   tight orbits, as demonstrated by the recent discovery of CoRoT-7b \citep{Leger2009,Queloz2009}.
   
   This paper reports the discovery and characterization of CoRoT-8b, a hot and dense small
   Saturn-class exoplanet in a 6.2-day orbit around a K1 dwarf star. We start by describing
   CoRoT LC analysis and transit parameter measurements. We then report ground-based
   follow-up observations and derive the full orbital and physical solution. After a look
   at transit timing variations and differential limbdarkening, we close this paper with
   a discussion about CoRoT-8b's possible composition.
   

\section{CoRoT-8 photometric observations} \label{sec:photom}

   CoRoT performs photometric measurements by onboard summation of all the photo-electrons
   collected during unit exposures of 32~s inside a predefined zone of the CCD matrix called
   a photometric mask. At the beginning of a run, every target star receives a mask whose
   shape is optimized given the star magnitude and environment. Mask sizes, ranging from
   about $4 \times 4$ to $15 \times 10$ pixels or $9 \times 9$ to $23 \times 35$ arcsec
   on the sky, are fairly large in order to accomodate CoRoT defocused point-spread functions
   (PSFs). On the one hand, defocusing improves the photometric precision and reduces the sensitivity
   to telescope jitter, but on the other, it increases the probability of PSF overlaps that
   may bias the photometry (if left uncorrected) or cause transit false alarms (if a neighbor
   is an eclipsing binary). In the exoplanet channel, PSFs are not only defocused but also
   slightly dispersed by a biprism. Thus, for a subset of stars, CoRoT delivers three LCs instead
   of one by division of their masks into three submasks along the dispersion direction.
   These ``color LCs'' are referred to as the red, green, and blue LCs, regardless of their
   true spectral content (and with no link whatsoever to any usual photometric system).   
   
   CoRoT-8 is one of the 11\,408 target stars observed by CoRoT during the first long run in
   constellation \textit{Aquila} (LRc01 for short) from May 16, 2007 to October 5, 2007.
   \cite{Cabrera2009} review all planetary transit candidates for this field, including CoRoT-8,
   which is referred to as LRc01 E2 1145-CHR-0101086161. This star belongs to the subset of 3719
   stars for which color LCs are available (CHR stands for chromatic).
   The standard CoRoT pipeline (version 2.1 at this date) produces 4 LCs in this case: one for
   every color and a ``white'' LC derived from the sum of colors. However, only the white LC receives
   proper telescope jitter noise correction, which is the reason color LCs have not been
   extensively studied in the previous papers of this series. For this study, we processed
   the white LC further by removing systematics effects as prescribed by \cite{Mazeh2009}, and
   we applied a generalization of this treatment to the colors (Mazeh et al., in preparation). 
   
\begin{figure}
      \centering
      \includegraphics[width=9cm]{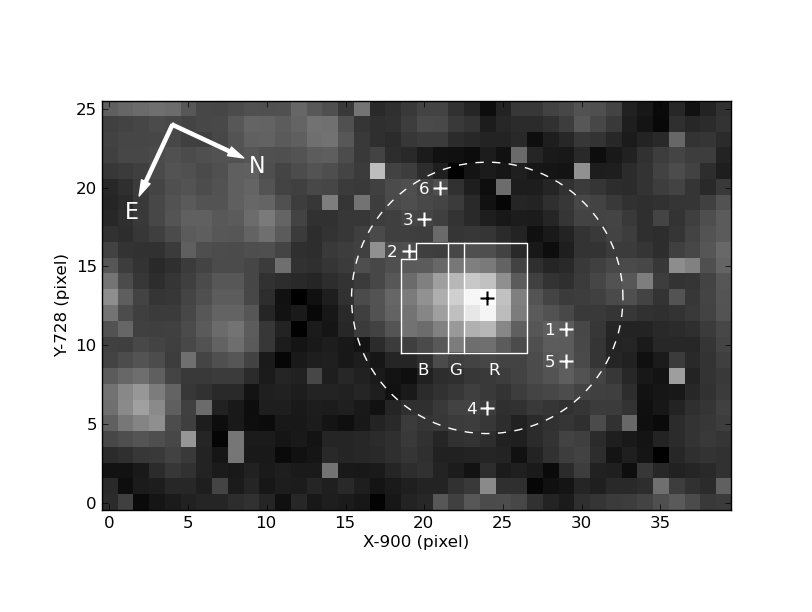}
      \caption{Long exposure taken at the beginning of LRc01 observation showing CoRoT-8 and its
      environment. The mask chosen for CoRoT-8 (solid white line) contains a total of 55 pixels.
      The letters B, G, and R indicate the positions of the blue (20 pix.), green (7 pix.),
      and red (28 pix.) submasks, respectively. The mask size on the sky is $18.6\arcsec \times 16.2\arcsec$
      (individual pixel size is $2.32\arcsec$). The white crosses indicate the positions
      of the six closest contaminants that lie inside a $40\arcsec$-diameter circle (dashed white
      line) centered on CoRoT-8. $X$ and $Y$ are positions on CCD E2.}
      \label{fig:mask1}
   \end{figure}

\begin{figure}
      \centering
      \includegraphics[width=9cm]{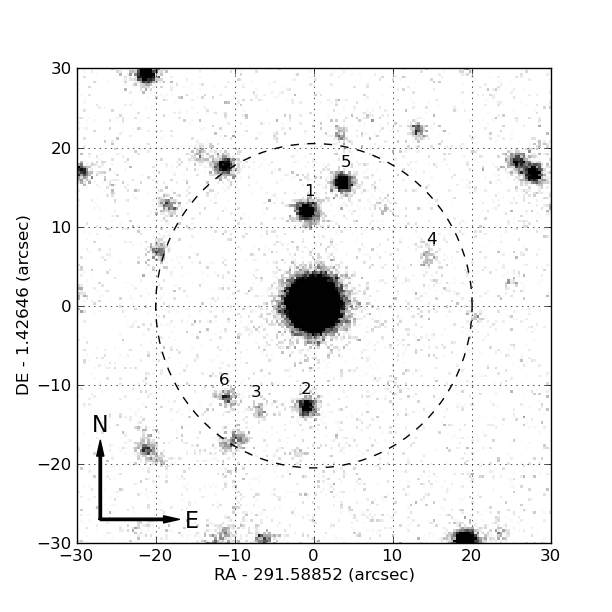}
      \caption{Image taken in the $r'$ filter with the WFC at the 2.5-m INT showing CoRoT-8 and
      its environment (seeing is $\sim 2\arcsec$).
      The six closest contaminants lying inside a $40\arcsec$-diameter circle (dashed black
      line) are labeled 1 to 6 according to increasing angular distances from CoRoT-8.}
      \label{fig:contaminants}
   \end{figure}

\begin{figure}
      \centering
      \includegraphics[width=9cm]{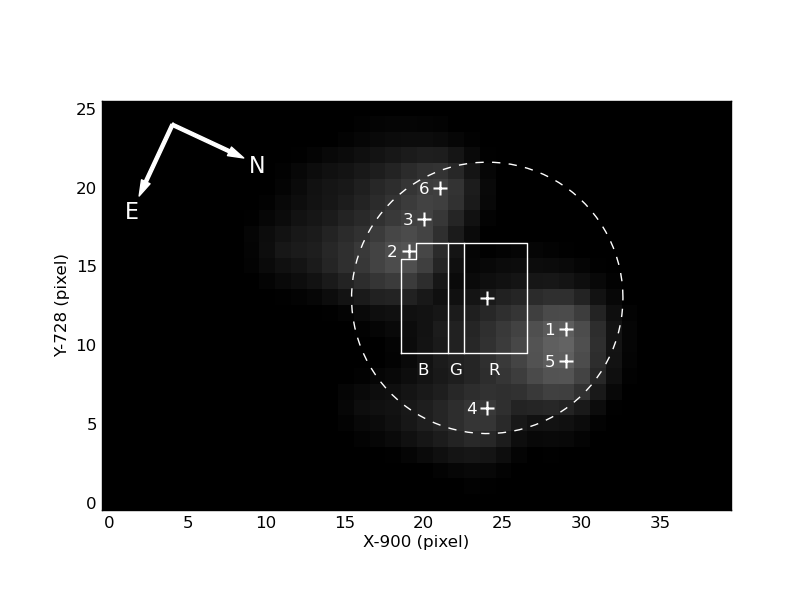}
      \caption{Computed analog of Fig.~\ref{fig:mask1} showing CoRoT-8's mask and
      its 6 closest contaminants using the same gray scale as Fig.~\ref{fig:mask1}.}
      \label{fig:mask2}
   \end{figure}

   At this stage, none of these LCs is corrected for the fluxes of neighboring stars that leak
   into the target star mask. Exodat, the CoRoT entry database \citep{Deleuil2009}, provides the
   necessary information to remove the contributions from these contaminants. An estimate of the
   contamination for the white LC is readily available (1.6\,\% in the case of CoRoT-8) but is
   computed with a generic mask prior to begining-of-run mask affectation. To get a
   more precise estimate of the contamination for the white, as well as nonexisting estimates
   for the colors, we superimposed inflight measured PSFs at the locations of the six main
   contaminants scaled with respect to their Sloan-Gunn $r'$ magnitudes (Table~\ref{tab:exodat}), then
   integrated their fluxes inside CoRoT-8's mask and submasks. Figure~\ref{fig:mask1} displays
   the long exposure taken at the begining of LRc01, CoRoT-8's mask, and the locations of the contaminants.
   For orientation, Fig.~\ref{fig:contaminants} is an image of the neighborhood of CoRoT-8 taken
   in 2004 with the Wide Field Camera (WFC) at the 2.5-m Isaac Newton Telescope (INT) at La Palma.
   Figure~\ref{fig:mask2} is a computed version of Fig.~\ref{fig:mask1} that ignores everything but CoRoT-8's
   and contaminants. This calculation yields new contamination estimates of 0.9\,\% for the white, 2.4\,\%
   for the blue, 0.2\,\% for the green, and 0.7\,\% for the red. In the following, LCs are corrected using
   these values.

   \begin{table}
	  \caption{Coordinates and magnitudes of CoRoT-8 and its six main
	  contaminants.}
      \label{tab:exodat}
      \centering
      \begin{tabular}{cccccccc}
      \hline\hline
      Star & RA (deg) & DE (deg) & $X$ & $Y$ & $V$ & $r'$ \\
      \hline
      CoRoT-8 & 291.58852 & 1.42646 & 924 & 741 & 14.8 & 14.3 \\
      Cont. \#1 & 291.58839 & 1.42976 & 929 & 739 & 18.8 & 18.3 \\
      Cont. \#2 & 291.58826 & 1.42280 & 919 & 744 & 19.1 & 18.8 \\
      Cont. \#3 & 291.58652 & 1.42269 & 920 & 746 & 20.7 & 20.1 \\
      Cont. \#4 & 291.59267 & 1.42809 & 924 & 734 & 20.7 & 20.1 \\
      Cont. \#5 & 291.58967 & 1.43077 & 929 & 737 & 19.0 & 18.6 \\
      Cont. \#6 & 291.58537 & 1.42313 & 921 & 748 & 20.3 & 19.7 \\
      \hline
      \end{tabular}
      \tablefoot{$X$ and $Y$ are positions on CCD E2 in pixels. $V$ and $r'$
	  are the magnitudes in the Harris $V$ and Sloan-Gunn $r'$ filters.}
   \end{table}
   
   CoRoT-8's 142.1-day long LCs consist of a first section of 11\,496 measurements with 512-s
   sampling period (sixteen 32-s unit exposures coadded onboard) and lasting 68.3 days. On
   the basis of this first section, the transit search pipeline called \emph{Alarm Mode} \citep{Surace2008} that
   continuously scrutinizes CoRoT LCs during a run, detected a transit-like signal and stopped
   unit-exposure onboard coaddition so that the remainder of this promising LC (198\,752
   measurements, 73.7 days) could benefit from a finer sampling period of 32~s. This action
   also triggered the beginning of ground-based follow-up operations: on-off transit
   photometry with the 1.2-m telescope at Observatoire de Haute-Provence (OHP) in France showed
   that the transit signal could not originate in any of the contaminants, so radial velocity
   (RV) spectrometry described in Sect.~\ref{sec:rv} was also started.
   
\begin{figure*}
      \centering
      \includegraphics[width=18cm]{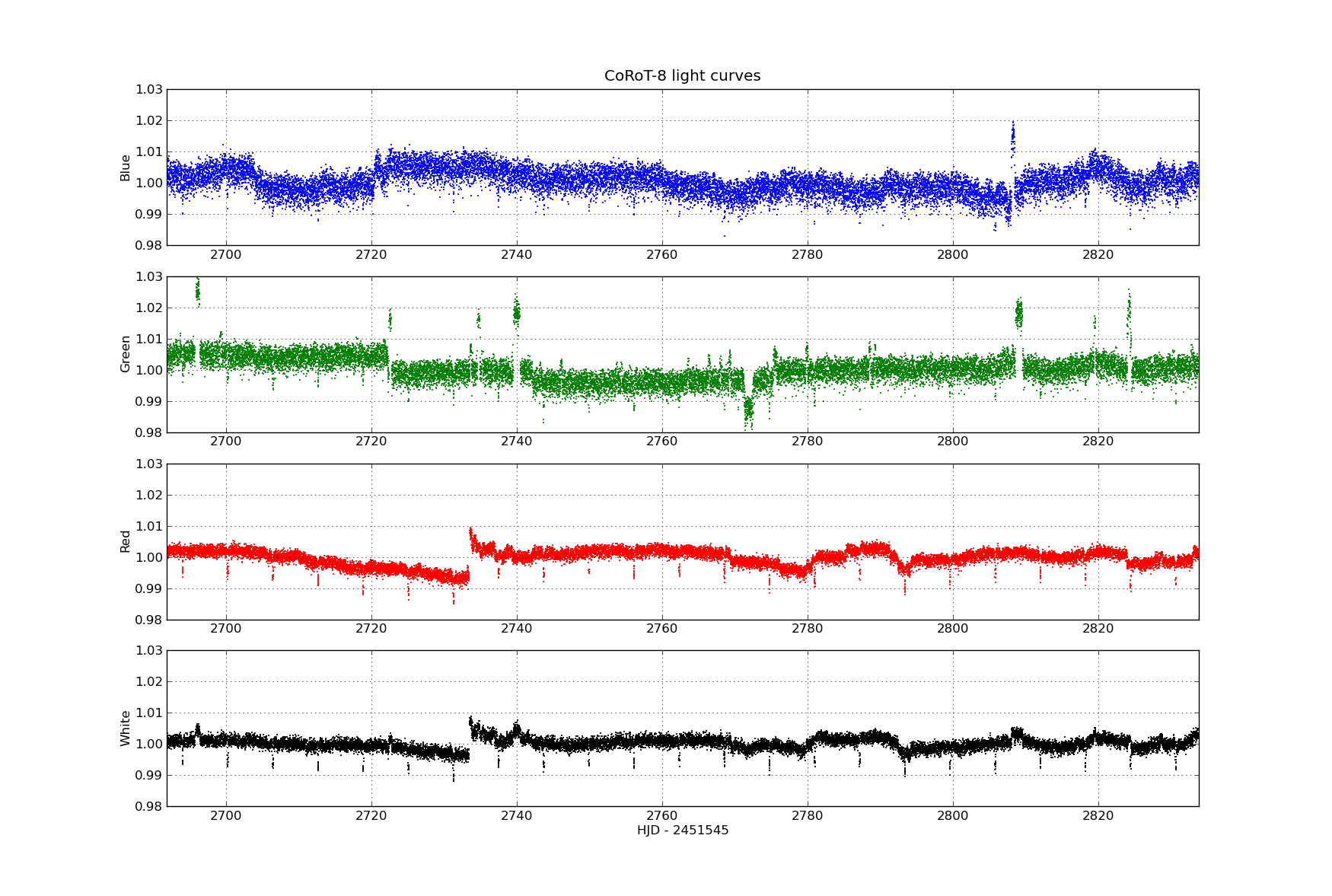}
      \caption{CoRoT-8's normalized light curves vs. time in Heliocentric Julian Date (HJD)
      minus 2\,451\,545 (January 1$^\mathrm{st}$, 2000). Sampling period is 512~s. The actual
      median fluxes are 14\,724, 15\,677, 80\,528, and $110\,408\,\mathrm{e}^-/32\,\mathrm{s}$
      for the blue, green, red, and white LCs, respectively.}
      \label{fig:lightcurves}
   \end{figure*}

   Figure~\ref{fig:lightcurves} shows the fully processed LCs (including systematics removal and
   contamination correction) rebinned to a uniform 512-s sampling period. The blue, green, and
   red fluxes amount to about 13\,\%, 14\,\%, and 73\,\%, respectively, of the total flux
   collected inside the photometric mask. Twenty-three transits with relative depths of about
   1\,\% can easily be seen by eye in the white and the red, are more difficult to see in the green,
   and are barely visible in the blue. Discontinuities are caused by energetic particles (mostly protons)
   from the Van Allen radiation belts that hit CoRoT when it crosses the South-Atlantic Anomaly
   \citep[SAA,][]{Pinheiro2008}.
   They differ from one color to the next depending on the locations of the pixels impacted by SAA
   protons. The green LC seems to be more polluted than the other two. Overall, these LCs do
   not show obvious signs of stellar activity, in contrast to CoRoT-7, for instance, where quasi-periodic
   modulations could be used by \cite{Leger2009} to estimate the stellar rotation period.
      
\section{White transit signal modeling} \label{sec:model}

   We begin by measuring the transit signal properties by
   least-square fitting to the white LC of a periodic trapezoidal model with 5 parameters:
   the period $P$, the ingress start time $T_1$, the ingress duration $T_{12}$ (taken as
   equal to the egress duration $T_{34}$), the bottom-part duration $T_{23}$, and the
   relative depth $\delta$.
   
   We obtain first estimates of $P$, $T_1$, the full transit duration $T_{14}$, and $\delta$
   by maximizing the cross-correlation with a periodic square-shape transit as explained by
   \cite{Borde2007}. The white LC is then chopped in 23 transit-centered segments lasting
   $3 \times T_{14}$. We normalize each segment and correct for local baseline flux
   variations by dividing by a parabola fitted to the out-of-transit flux variations. Then, we
   proceed to the least-square fitting of the 5-parameter trapezoid model described above,
   assuming measurement errors to be identical and equal to $8.7\,10^{-4}$, the standard
   deviation of out-of-transit measurements. We find that $P = 6.21229 \pm 0.00003$~days,
   $T_1 = 2\,454\,238.9743 \pm 0.0004$~HJD, $\delta = (6.51 \pm 0.07)\times 10^{-3}$,
   $T_{12} = 41.9 \pm 0.8$~min, and $T_{23} = 80.4 \pm 0.8$~min with $\chi^2_\mathrm{r} = 1.4$.
   
   We keep the determined period and ingress start time for what follows, and we go on
   by fitting the white LC with a classical model that describes the transit of an
   opaque black disk (the planet) in front of a limb-darkened (LD)
   luminous disk (the star). The normalized measured flux is $f=1-F_p/F_\star$,
   where $F_p$ is the flux occulted by the planet and $F_\star$ the flux emitted
   by the star. If $I(x)$ is the normalized centro-symmetric LD profile of the star
   and $x$ the normalized distance to the star center, then the total flux emitted
   by the star is $F_\star = 2 \pi \int_0^1 I(x)\,x\,\mathrm{d} x$.
   Denoting $r=R_p/R_\star$ as the ratio of the planet radius to the star radius,
   the flux $F_p$ occulted by the planet at linear distance $\rho$ from the star
   center can be computed via
   \begin{equation} \label{eq:Fp1}
      2 \int_{\rho-r}^{\rho+r} I(x)\,\arccos \left(\frac{x^2+\rho^2-r^2}{2 \rho x}
      \right) \,x\,\mathrm{d} x
   \end{equation}
   if $r \le \rho < 1+r$, and with
   \begin{equation} \label{eq:Fp2}
      2\pi \int_0^{r-\rho} I(x)\,x\,\mathrm{d} x + 2 \int_{r-\rho}^{r+\rho} I(x)
      \,\arccos \left(\frac{x^2+\rho^2-r^2}{2 \rho x} \right) \,x\,\mathrm{d} x
   \end{equation}
   if $0 \le \rho < r$. Besides,
   \begin{equation} \label{eq:rho}
      \rho = \frac{a}{R_\star} \sqrt{\cos^2(i) \cos^2 \left( \frac{2\pi
      \,(t-T_0)}{P} \right) + \sin^2 \left( \frac{2\pi\,(t-T_0)}{P} \right)},
   \end{equation}
   where $a$ is the semi-major axis, $i$ the orbit inclination, $P$ the orbital
   period, and $T_0$ the central time of the first transit. To consider
   the effect of signal integration over 512-s exposures, our model is computed every
   8 s, then numerically integrated over 64 points. In the course of
   our work on CoRoT planets, this set of equations and its numerical
   implementation have been checked against other published approaches
   \citep{Mandel2002,Gimenez2006}.
   
   From Eqs.~\ref{eq:Fp1}--\ref{eq:rho}, we see that this model's parameters are
   the radius ratio $R_p/R_\star$, the scaled semi-major axis $a/R_\star$, the
   orbital inclination $i$, the transit central time $T_0$, and the orbital period
   $P$, plus usually 1--4 LD parameters depending on the form given to $I(x)$.
   In the following, we do not fit $T_0$ and $P$, but instead take the values obtained
   with the trapezoid model, and use the quadratic LD law $I(\mu)=1-u_1\,(1-\mu)-u_2\,(1-\mu)^2$
   (where the fitted LD parameters are actually $u_+=u_1+u_2$ and $u_-=u_1-u_2$).
   Therefore, we have a total of 5 parameters to determine.
   
\begin{figure}
      \centering
      \includegraphics[width=9cm]{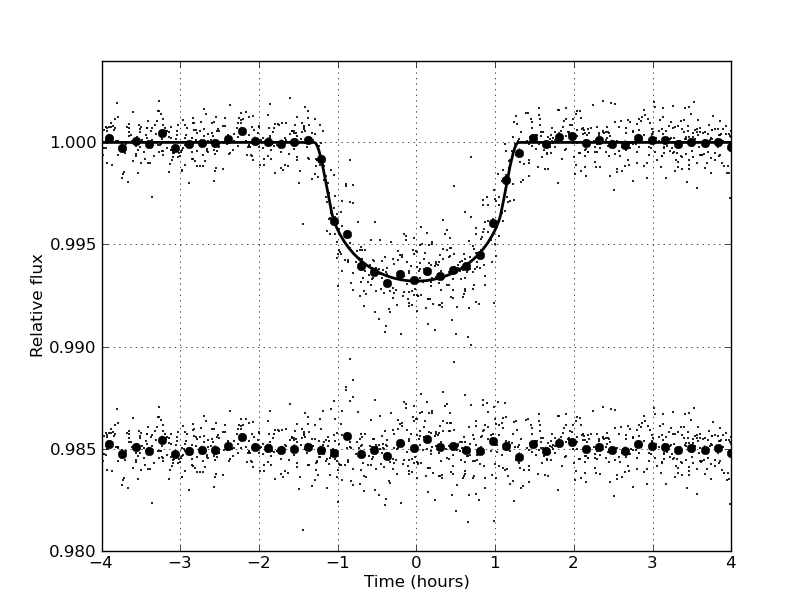}
      \caption{CoRoT-8b's best transit model for the white LC (solid line) with
      individual measurements (small dots) and averaged measurements inside 10-min
      bins (big dots). Fit residuals are displayed around 0.985.}
      \label{fig:bestmodel}
   \end{figure}

   Parameter optimization, done by way of a genetic algorithm called \emph{Harmony
   Search} \citep{Geem2001}, leads to $R_p/R_\star = 0.075 \pm 0.001$,
   $a/R_\star = 17.6 \pm 0.4$, $i = 88.4 \pm 0.1$\degr, $u_1 = 0.70 \pm 0.09$, and
   $u_2 = 0.14 \pm 0.09$ with $\chi^2_\mathrm{r} = 1.4$ (Fig.~\ref{fig:bestmodel}).
   Individual parameter errors were found by increasing $\chi^2$ from
   $\chi^2_\mathrm{min}$ to $\chi^2_\mathrm{min}+1$ as prescribed by \cite{Press1997}.
   From the scaled semi-major axis and the period, we obtain (via Kepler's third law)
   $M_\star^{1/3}/R_\star = 1.24 \pm 0.03$ in solar units, hence a stellar density
   $\rho_\star = 2.7 \pm 0.1$~\gcm3 (see Table~\ref{tab:param} for a summary).
   To compute the planet radius, we now need to discuss the star properties.

\section{CoRoT-8 stellar parameters} \label{sec:star}

   CoRoT-8 was observed with the UVES spectrograph on the VLT (program 081.C-0413) on September 2 and 3,
   2008. We used the  Dic1 mode (346+580) and a slit width of $0.5\arcsec$ that provides a
   resolving power of $\sim 65\,000$. The total exposure time was 4760~s, divided into two
   exposures. The data were reduced using our own reduction software to optimize the
   spectrum extraction. The spectra were divided by the blaze function and put at rest by subtracting
   the systemic radial velocity. We obtained a signal-to-noise ratio ($S/N$) of about 170 per resolution
   element at 550~nm.

   The fundamental parameters of the star were first estimated with this spectrum. Later, we took
   advantage of the series of HARPS spectra to refine the estimates of $v\sin i$ and the micro and
   macro-turbulent velocities, which all benefit from the much higher spectral resolution of HARPS.
   To that purpose, among the HARPS spectra collected for the radial velocity analysis
   (Sect.~\ref{sec:rv}), we selected those that were not contaminated by Moon light. Nine spectra
   were finally co-added, yielding an average spectrum with $S/N \sim 150$ per resolution
   element at 550~nm.

   The star is a slow rotator since the rotational broadening is lower than the spectral resolution
   of HARPS. Using the synthesis of a set of isolated lines, we determined a projected rotational
   velocity of $v \sin i = 2 \pm 1$~\kms, and a macroturbulent velocity of
   $v_\mathrm{mac} = 2 \pm 1$~\kms. For the spectroscopic analysis, we followed the same methodology
   as for the previous CoRoT planets since CoRoT-3b. The analysis was performed using two spectral
   synthesis packages, VWA \citep{Bruntt2002} and SME \citep{Valenti1996,Valenti2005} as described
   in details by \cite{Bruntt2009}.
   
   For such a late type star that is metal rich, the continuum placement could be a serious issue
   at wavelengths shorter than 550~nm owing to molecular bands and the high level of blending of weak
   metal lines. We therefore concentrated our spectroscopic analysis on well-defined spectral lines at longer
   wavelengths. We used the \ion{Ca}{i} lines at 612.2, 616.2, and 643.9~nm as surface gravity
   diagnostics instead of the usual \ion{Mg}{i}b or \ion{Na}{i} lines. This overcomes the issue of
   continuum placement for such broad lines with extended wings when observed with an echelle
   spectrum. To provide an additional check of our effective temperature determination,
   we compared the $\mathrm{H}\alpha$ line profile outside of the core to theoretical profiles,
   following the method recommended in Sect.~3.3 of \cite{Bruntt2004}. The result agrees with our
   estimates obtained independently with VWA and SME. Our final atmospheric parameters,
   $T_\mathrm{eff}$, $\log g$, and global metallicity estimates are given in Table~\ref{tab:param}.
   The results from the detailed elemental abundances analysis are summarized in Table~\ref{tab:abund}.
   These correspond to a slowly rotating K1 main-sequence star with noticeable metal enrichment.

   To estimate the mass and the radius of the host star, we combined the constraints from the
   spectroscopic parameters ($T_\mathrm{eff}$ and [M/H]) and $M_\star^{1/3}/R_\star$,
   the proxy for stellar density derived from the transit physical model (see Sect.~\ref{sec:model}
   and Table~\ref{tab:param}). The spectroscopic analysis of our spectra gave no evidence of youth:
   the \ion{Li}{i} line at 670.78~nm is absent and the star's spectra display no hint of chromospheric
   activity. This is consistent with the low rotational velocity we measured. Using STAREVOL stellar
   evolutionary tracks \cite[Palacios, \emph{private communication}]{Siess2006}, we thus excluded
   evolutionary tracks corresponding to pre-main sequence stages and assumed that the star is on
   the main sequence. In this way, we obtained a stellar mass of $M_\star = 0.88 \pm 0.04 \:
   \mathrm{M}_\odot$ and furthermore inferred a stellar radius of $R_\star = 0.77 \pm 0.02 \:
   \mathrm{R}_\odot$. The corresponding surface gravity is $4.61\pm 0.07$ in good agreement with
   the spectroscopic value of $\log g = 4.58 \pm 0.08$.
   
   Concerning limb-darkening, we also note fair agreement between measured values reported
   in Sect.~\ref{sec:model}, $u_1 = 0.70 \pm 0.09$ and $u_2 = 0.14 \pm 0.09$, and theoretical
   estimates computed for the CoRoT bandpasses by \cite{Sing2010}: $u_1 = 0.59$ and $u_1 = 0.12$
   (for $T_\mathrm{eff}=5000$~K, $\log g=4.5$, and $[\mathrm{M}/\mathrm{H}]$=0.3). Achieving
   consistency in both surface gravity and limb-darkening is a good indication that the
   determined orbital and physical parameters are reliable. Therefore, we can safely derive a
   planetary radius of $0.57 \pm 0.02 \: \mathrm{R}_\mathrm{J}$, which is 1.7 times the radius
   of Neptune and 0.7 times that of Saturn.

   \begin{table}
      \caption{Abundances of the main elements in CoRoT-8.}
      \label{tab:abund}
      \centering
      \begin{tabular}{ccc}
      \hline\hline
      Element & $\mathrm{\textbf{[X/H]}}$ & number \\
              &                  & of lines \\
      \hline
      {Mg \sc   i} &  0.16  (0.16)  &   3 \\
      {Ca \sc   i} &  0.31  (0.05)  &   4 \\
      {Ti \sc   i} &  0.26  (0.05)  &  14 \\
      {V  \sc   i} &  0.49  (0.07)  &  16 \\
      {Fe \sc   i} &  0.31  (0.05)  &  96 \\
      {Fe \sc  ii} &  0.32  (0.06)  &   8 \\
      {Co \sc   i} &  0.30  (0.07)  &   6 \\
      {Ni \sc   i} &  0.34  (0.05)  &  30 \\
      {Si \sc   i} &  0.37  (0.05)  &  18 \\
      \hline
      \end{tabular}
   \end{table}


\section{Radial velocity spectrometry} \label{sec:rv}
   
	We performed RV observations of CoRoT-8 with the SOPHIE 
	spectrograph \citep{Perruchot2008,Bouchy2009a} at the 1.93-m telescope located at OHP,
	and the HARPS spectrograph \citep{Mayor2003} at the 3.6-m ESO telescope at La Silla
	Observatory (Chile). These two instruments are cross-dispersed, fiber-fed, echelle
	spectrographs dedicated to high-precision Fizeau-Doppler measurements. SOPHIE was used
	in its high-efficiency mode (spectral resolution of 40\,000). HARPS and SOPHIE were both
	used in the obj\_AB observing mode, where the second fiber was made available for
	monitoring the background Moon light at the cost of no simultaneous Th-Ar lamp calibration.
	This is made possible thanks to the remarkable intrinsic stability of these spectrographs.
	Indeed, the contribution of instrumental drift to the global RV errors was always negligible
	with respect to the photon noise error. HARPS and SOPHIE data were reduced with the standard
	pipeline based on the work by \cite{Baranne1996}, \cite{Pepe2002}, and \cite{Bouchy2009a}.
	In the case of CoRoT-8, radial velocities were derived by cross-correlating the acquired
	spectra with a numerical K5 mask.
	
	The first two measurements on CoRoT-8 were made with SOPHIE on August 2007 and showed
	no significant RV variations at the level of 50~\ms. As such, they were compatible
	with a planetary companion lighter than 1 {\Mjup}, so it called for more observations.
	Nine more measurements were obtained with HARPS between the end of July 2008 and the
	end of September 2008 (ESO program 081.C-0388). Although they yielded RV variations in phase
	with the ephemerides computed from the CoRoT-8 light curve, they did not allow by themselves
	properly constraining of the orbit, and above all completely ruling out the presence of
	a blended binary (a background eclipsing binary not angularly separated from
    CoRoT-8 by the 3.6-m ESO telescope). Therefore, ten additional HARPS measurements were
    acquired to secure the planetary origin of the CoRoT transits from June to August 2009
    (ESO program 083.C-0186).
	
\begin{figure}
      \centering
      \includegraphics[width=9cm]{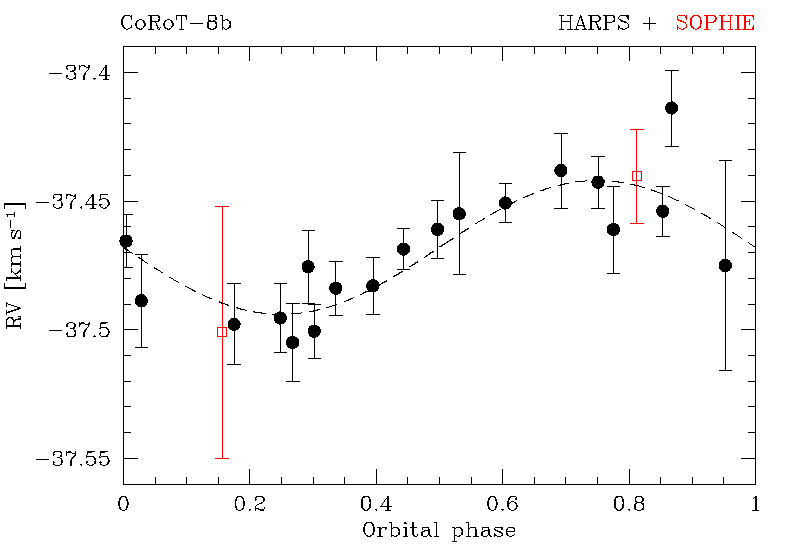}
      \caption{Period-folded radial velocities of CoRoT-8. The black dots and open red squares
      respectively correspond to HARPS and SOPHIE measurements.}
      \label{fig:rv}
   \end{figure}

	All RV measurements (Table~\ref{tab:rv}) are displayed in Fig.~\ref{fig:rv}.
	The SOPHIE data were shifted by 105~\ms for this plot.
	Several spectra of the third campaign were slightly contaminated by Moon light. We attempted
	to correct for the Moon signal, but it only introduced more noise and did not significantly 
	affect the target RV values. (The Moon RV was different enough from that of the target.)
	We concluded from this study that Moon contamination could be safely ignored in our case.  

   The period-folded radial velocities in Fig.~\ref{fig:rv} present clear variations in phase with
   ephemerides derived from CoRoT transits, in agreement with what would be expected for the reflex
   motion due to a planetary companion. We fitted RV data with a Keplerian model assuming a circular
   orbit and found a systemic RV of $-37.468 \pm 0.003$~\kms and an RV semi-amplitude of $26\pm4$~\ms
   with $\chi^2_r=0.8$ (showing that the standard deviation of the residuals equal to $11.8$~\ms has
   the same order of magnitude as the individual RV errors).

\begin{figure}
      \centering
      \includegraphics[width=9cm]{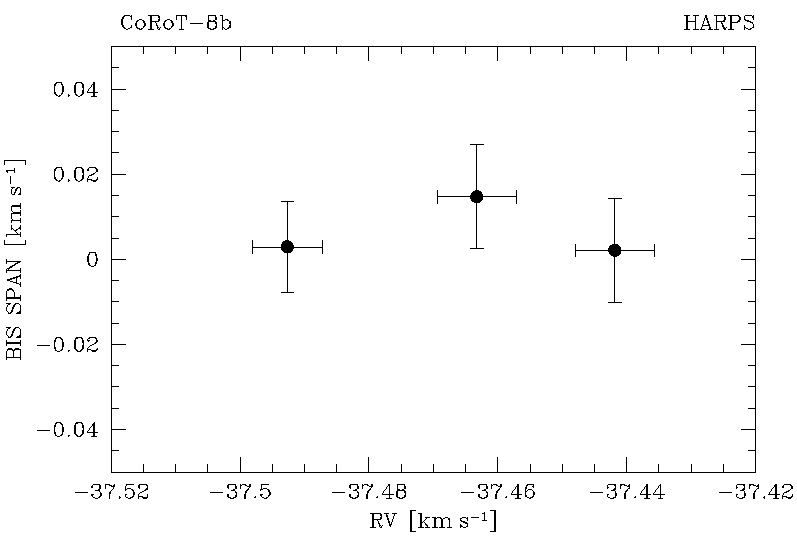}
      \caption{Bisector span vs. radial velocity. The three points shown are averages obtained
      from six, five, and five cross-correlation functions computed in orbital phase intervals
      0.15--0.35, 0.40--0.61, and 0.65--0.86, respectively.}
      \label{fig:bisector}
   \end{figure}

   To examine the possibility that the radial velocity variation is caused by a blended binary,
   we followed the procedure described in \cite{Bouchy2009b}. It consists in checking the spectral
   line assymmetries, as well as the dependence of RV variations on the cross-correlation mask used
   to compute the cross-correlation function (CCF). Spectral line assymetries are revealed by the
   calculation of the bisector span of the CCF. To increase the signal-to-noise ratio, we
   first averaged CCFs around orbital phases of 0.25, 0.5, and 0.75: six, five, and five CCFs were thus
   averaged in orbital phase intervals 0.15--0.35, 0.40--0.61, and 0.65--0.86, respectively. Bisector
   spans computed on these three averaged CCFs show no significant correlation with the RV as can be
   seen in Fig.~\ref{fig:bisector}, since the $rms$ value of the bisector span variation is
   less by a factor of five than the radial-velocity semi-amplitude $K$. Furthermore, RV computed
   with different cross-correlation templates (for F0, G2, and K5 stars) do not correlate with the
   amplitude of RV.
   
   These two checks are common tools for discarding the blended binary scenario and confirming the planetary
   nature of transiting planets. However, to support this conclusion further, we also contemplated
   the two scenarios: 1) a triple system, and 2) a background eclipsing binary blended with
   CoRoT-8. In the first case, it was found that only a brown-dwarf companion orbiting an M dwarf
   could produce the same RV and bisector signals. However, such a system would have a relative
   transit depth twice as large as the one measured by CoRoT. In the second case, the velocity of
   CoRoT-8 and the systemic velocity of the background binary would have to be the same at the level
   of 1 km/s in order to give a symmetrical spectroscopic signal. This would be very unlikely when
   considering the dispersion of the velocity distribution in the Galaxy. Therefore, the most probable
   cause of the RV signal is a planetary companion to CoRoT-8.
   
   CoRoT-8b's mass (in \Mjup) can then be computed as $4.92\,10^{-3} \, K \, M_\star^{2/3} \, P^{1/3}$,
   where $K$ is in \ms, $M_\star$ in $\mathrm{M}_\odot$, and $P$ in days. We find $M_p = 0.22 \pm 0.03$~\Mjup,
   4.1 times the mass of Neptune and 0.7 times that of Saturn. Other stellar and planetary parameters
   can be found in Table~\ref{tab:param}.
   
   Finally, no significant offset was found between the two sets of HARPS measurements separated by
   one year, thus excluding the presence of an additional massive giant planet at a few AUs in the
   system.

   \begin{table}
      \caption{Radial velocity measurements on CoRoT-8 obtained by
      HARPS and SOPHIE spectrometers.}            
      \centering                          
      \begin{tabular}{ccccc}       
      \hline\hline                 
      BJD\tablefootmark{a} & RV & 1-$\sigma$ error & exp. time & S/N p. pix. \\
      $-2\,400\,000$ & (\kms) & (\kms) & (s) &  (at 550~nm) \\
      \hline                        
      \multicolumn{5}{c}{SOPHIE} \\
      \hline
      54314.40776 & $-37.6060$ & 0.0490 & 3600 & 12.15 \\
      54318.41015 & $-37.5454$ & 0.0184 & 3600 & 17.00 \\
      \hline 
      \multicolumn{5}{c}{HARPS} \\
      \hline
      54675.69652 &	$-37.5050$ & 0.0151 & 2700 & 9.20 \\
      54678.66557 & $-37.4427$ & 0.0101 & 2700 & 12.20 \\  
      54679.73180 &	$-37.4751$ & 0.0407 & 2700 & 4.60 \\   
      54682.68436 &	$-37.4830$ & 0.0112 & 2700 & 11.20 \\  
      54700.57459 &	$-37.4756$ & 0.0140 & 2700 & 8.90 \\  
      54703.59565 &	$-37.4611$ & 0.0169 & 2700 & 8.40 \\  
      54732.56202 &	$-37.4687$ & 0.0080 & 3600 & 14.50 \\  
      54733.58576 &	$-37.4508$ & 0.0077 & 3600 & 14.80 \\  
      54737.58497 &	$-37.4955$ & 0.0134 & 3600 & 10.30 \\  
      54987.78368 &	$-37.4550$ & 0.0237 & 3600 & 7.82 \\  
      54988.84456 &	$-37.4482$ & 0.0146 & 3600 & 10.92 \\  
      54989.83528 &	$-37.4540$ & 0.0098 & 3600 & 14.77 \\  
      54991.85115 &	$-37.4979$ & 0.0157 & 3600 & 10.74 \\  
      54992.80807 &	$-37.4839$ & 0.0106 & 3600 & 13.68 \\  
      54993.87346 &	$-37.4610$ & 0.0113 & 3600 & 13.30 \\  
      54998.85102 &	$-37.5006$ & 0.0106 & 3600 & 12.20 \\  
      55021.82951 &	$-37.4656$ & 0.0102 & 3600 & 14.22 \\  
      55045.74089 &	$-37.4139$ & 0.0148 & 3600 & 10.95 \\  
      55046.73467 &	$-37.4888$ & 0.0179 & 3600 & 9.47 \\  
      \hline                                   
      \end{tabular}
      \tablefoottext{a}{Barycentric Julian Date}
      \label{tab:rv}      
   \end{table}


\section{Search for another body by transit timing} \label{sec:ttv}

   In this section, we look for transit timing variations (TTVs) in the white LC that may be caused by
   a gravitational perturber \citep[e.g.,][]{Sartoretti1999,Schneider2003}. To do this,
   we fit a trapezoid to each transit separately in which the ingress time, $T_1$, is the
   only free parameter and other parameters are held constant at the values reported at the
   beginning of Sect~\ref{sec:model}. 
   
   \begin{figure}
      \centering
      \includegraphics[width=9cm]{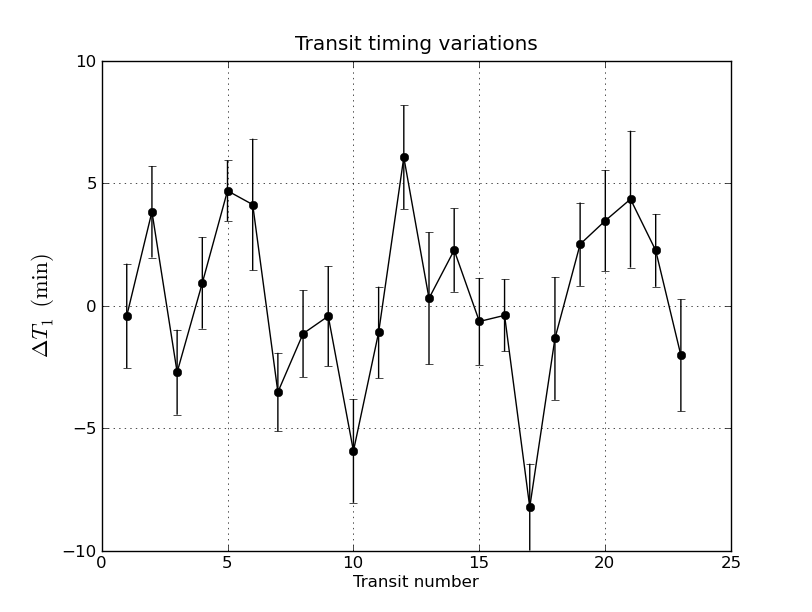}
      \caption{Transit-to-transit variation of the date of first contact as a function of the
      transit number.}
      \label{fig:ttv}
   \end{figure}
   
   We find variations that seem to be statistically significant (Fig.~\ref{fig:ttv}) and whose
   autocorrelation curve peaks at $7P = 43.5$~days. However, we cannot be assured that these
   variations are gravitational in origin. To see whether they might be caused by stellar
   activity in the LC, we performed a dedicated analysis of the LC long-term variations to
   measure the stellar rotation period. For this
   purpose, we selected the red LC as the best compromise between the photometric precision
   and the number of discontinuities (Fig.~\ref{fig:lightcurves}). We corrected this LC for two
   main discontinuities around dates 2733 and 2824 and removed data points inside the transits.
   The autocorrelation of the resulting LC has a weak local maximum around 20 days ($20 \pm 5$ days).
   If we cut this 142-day LC into three thirds lasting about 47 days each, we again find weak local
   maxima around 20 days for the first and the third thirds, but not for the second one. Although uncertain,
   this tentative signal around 20 days might still reflect the rotation period of the star. Indeed, if
   we combine the measurements for $v \sin i$, $R_\star$, and $i$, we derive a loosely constrained
   stellar rotation period of $20 \pm 10$ days. Now, activity-induced TTVs may be periodic if the
   planet transits repeatedly above the same stellar surface features, which implies that the ratio of the stellar
   rotation period to the planet orbital period be a rational number. If $7P/2 = 21.7$~days is closer to
   the true stellar rotation period than $3P = 18.6$~days, this may explain why there seem to be
   a TTV signal at about twice the stellar rotation period.
   
\begin{figure}
      \centering
      \includegraphics[width=9cm]{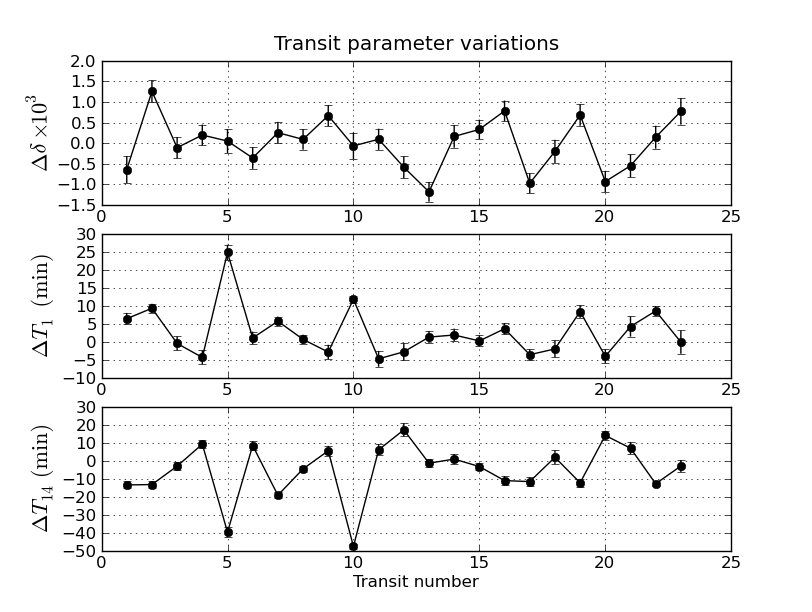}
      \caption{Transit-to-transit variations in the relative depth, date of first contact, and
      first to fourth contact duration as a function of the transit number.}
      \label{fig:tpv}
   \end{figure}

   We also explored transit-to-transit variations of $\delta$, $T_1$, and $T_{14}$
   at the same time by fitting trapezoids to individual transits where $\delta$, $T_1$, $T_{12}$,
   and $T_{23}$ are now all free parameters. Figure~\ref{fig:tpv} displays the resulting variations,
   denoted $\Delta \delta$, $\Delta T_1$, and $\Delta T_{14}$, as a function of the transit number.
   The last two variations are obviously highly anti-correlated (with a correlation coefficient of
   $-0.82$) showing that variations in date actually reflect variations in duration. As this should
   not be the case for a gravitational pertuber, we are inclined to believe that the TTVs we measure
   are more likely to be the effect of stellar noise.
   
\section{Differential limb-darkening} \label{sec:color}

   With its photometry in three colors, CoRoT makes it possible in principle to study differential
   stellar limb-darkening. However, if the photometric accuracy in the red is comparable to the
   white, it is not the case for the green and the blue where the collected flux is about five
   times lower than in the red (Sect.~\ref{sec:photom}). Indeed, the standard deviation of out-of-transit
   measurements is $8.0\,10^{-4}$ for the red, and jumps to $2.4\,10^{-3}$ for
   the blue and $2.1\,10^{-3}$ for the green. Moreover, satellite jitter noise and contamination
   by neighboring stars (Sect.~\ref{sec:photom}) increase the difficulty of this endeavor.
   Nevertheless, we make a first attempt here to study CoRoT-8b's transit in the CoRoT colors.
   
   As a start, we first used the same trapezoidal model as for the white but we fixed $P$, $T_1$,
   and $T_{14}$ at the values obtained with the white LC (as we do not expect variations here),
   and fit only for $\delta$ and $T_{12}$.
   We find that the transit ingress duration is longer in the blue (51~min) than in the green
   (46~min) and longer in the green than in the red (39~min), which is the expected effect of
   the more pronounced star limb-darkening at shorter wavelengths, and this seems encouraging.
   The blue transit ($(5.71 \pm 0.17)\times 10^{-3}$) is also slightly deeper than the green
   ($(5.14 \pm 0.14)\times 10^{-3}$), as expected, but is unfortunately not deeper than the red
   ($(5.71 \pm 0.17)\times 10^{-3}$), which suggests that color corrections might still not be
   accurate enough for CoRoT-8.

   Nevertheless, if we attempt to model the color transits with the white model from
   Sect.~\ref{sec:model} by holding constant $R_p/R_\star$, $a/R_\star$, and $i$, and letting
   $u_1$ and $u_2$ vary, we do measure limb-darkening parameters for the colors
   (Table~\ref{tab:limbdarkening}), but with a moderate fit quality in the green and the blue.
 
   \begin{table}
	  \caption{Comparison of limb-darkening parameters measured from the white and color LCs.}
      \label{tab:limbdarkening}
      \centering
      \begin{tabular}{ccc}
      \hline\hline
      LC    & $u_1$ & $u_2$ \\
      \hline
      white & $0.70 \pm 0.09$ & $0.14 \pm 0.09$ \\
      blue  & $0.88 \pm 0.04$ & $-0.04 \pm 0.04$ \\
      green & $0.68 \pm 0.16$ & $-0.25 \pm 0.16$ \\
      red   & $0.46 \pm 0.12$ & $0.26 \pm 0.12$ \\
      \hline
      \end{tabular}
   \end{table}
   
   It also worth reporting that, when $R_p/R_\star$, $a/R_\star$, and $i$ are included
   as free parameters in the modeling of the three colors taken separately, the obtained best-fit
   values for these parameters are all consistent with the white solution. This consistency
   check brings further evidence that the transiting object is a genuine exoplanet.      


\section{Discussion}

   \begin{table}
      \caption{Star and planet characteristics of the CoRoT-8 system.}                  
      \centering                         
      \begin{tabular}{l l}        
      \hline\hline
      CoRoT-ID & 0101086161 \\
      CoRoT-WinID & LRc01-E2-1145 \\
      GSC 2.3 & N29S083644 \\
      Coordinates (J2000) & 19:26:21.245 +01:25:35.55 \\
      Magnitudes B, V, r', i' & 16.10, 14.80, 14.27, 13.41 \\
      \multicolumn{2}{l}{\textbf{Results from light-curve analysis}} \\           
      Planet period, $P$                         & $6.21229 \pm 0.00003$ days \\
      Transit epoch, $T_0$                       & HJD $2\,454\,238.9743 \pm 0.0004$ \\
      Transit duration, $T_{14}$                 & $2.74 \pm 0.02$~h \\
      Transit relative depth, $\delta$           & $(6.51 \pm 0.07)\times 10^{-3}$ \\
      Radius ratio, $R_p/R_\star$                & $0.075 \pm 0.001$ \\
      Scaled semi-major axis, $a/R_\star$        & $17.6 \pm 0.4$ \\
      Orbital inclination, $i$                   & $88.4 \pm 0.1$\degr \\
      Impact parameter, $b$                      & $0.49 \pm 0.04$ \\
      Limb-darkening coefficients, $u_1$, $u_2$  & $0.70 \pm 0.09$, $0.14 \pm 0.09$ \\
      Stellar density, $\rho_\star$              & $2.7 \pm 0.1$~\gcm3 \\
      $M_\star^{1/3}/R_\star$                    & $1.24 \pm 0.03$ (solar units) \\
      \multicolumn{2}{l}{\textbf{Results from radial velocity observations}} \\
      Radial velocity semi-amplitude, $K$        & $26 \pm 4$~\ms \\
      Orbital eccentricity, $e$                  & 0 (fixed) \\
      Systemic radial velocity                   & $-37.468 \pm 0.003$~\kms \\
      %
      \multicolumn{2}{l}{\textbf{Results from spectral typing}} \\
      Spectral type                              & K1V \\
      Effective temperature, $T_\mathrm{eff}$    & $5080 \pm 80$~K \\
      Surface gravity, $\log g$                  & $4.58 \pm 0.08$ (cgs) \\
      Metallicity, $[\mathrm{M}/\mathrm{H}]$     & $0.3 \pm 0.1$  \\
      Micro-turbulent velocity, $v_\mathrm{mic}$ & $0.8 \pm 0.2$~km/s \\
      Macro-turbulent velocity, $v_\mathrm{mac}$ & $2 \pm 1$~km/s \\
      Rotational velocity, $v \sin i$            & $2 \pm 1$~\kms \\
      Star mass, $M_\star$                       & $0.88 \pm 0.04 \: \mathrm{M}_\odot$ \\
      Star radius, $R_\star$                     & $0.77 \pm 0.02 \: \mathrm{R}_\odot$ \\
      Star age                                   & $\le 3$~Ga \\
      Star distance                              & $380 \pm 30$~pc \\
      \multicolumn{2}{l}{\textbf{Absolute physical parameters from combined analysis}} \\
      Planet mass, $M_p$                         & $0.22 \pm 0.03$~\Mjup \\
      Planet radius, $R_p$                       & $0.57 \pm 0.02$~\Rjup \\
      Planet density, $\rho_p$                   & $1.6 \pm 0.1$~\gcm3  \\
      Planet orbital semi-major axis, $a$        & $0.063 \pm 0.001$~AU \\
      \hline                                  
      \end{tabular}
      \label{tab:param}      
   \end{table}

   With a transit depth of about 7~mmag, a transiting planet like CoRoT-8b is almost impossible
   to detect for ground-based photometric surveys that are not sensitive to giant planets smaller
   than Saturn: only HAT-P-11\,b \citep{Bakos2009} with a smaller transit depth (4.3~mmag) was
   detected by a ground-based photometric survey. CoRoT-7b \citep[0.3~mmag,][]{Leger2009} and
   Kepler-4b \citep[0.9~mmag,][]{Borucki2010} were both detected by space-based photometric surveys,
   whereas GJ\,436\,b \citep[0.6~mmag,][]{Gillon2007} and HD\,149026\,b \citep[2.6~mmag,][]{Sato2005}
   were both detected by RV surveys prior to their photometric observations.
   
   With the exception of CoRoT-10 (V\,=\,15.2, Bonomo et al., in preparation), CoRoT-8 (V\,=\,14.8) 
   is the faintest star harboring a planet found by the CoRoT survey. It is also relatively faint
   by RV standards, so that about 20 HARPS measurements (totalizing about 20 hours of telescope time)
   split over two campaigns were needed to definitively establish the planetary nature and 
   constrain the mass of CoRoT-8b within 14\,\%.
   
\begin{figure}
      \centering
      \includegraphics[width=9cm]{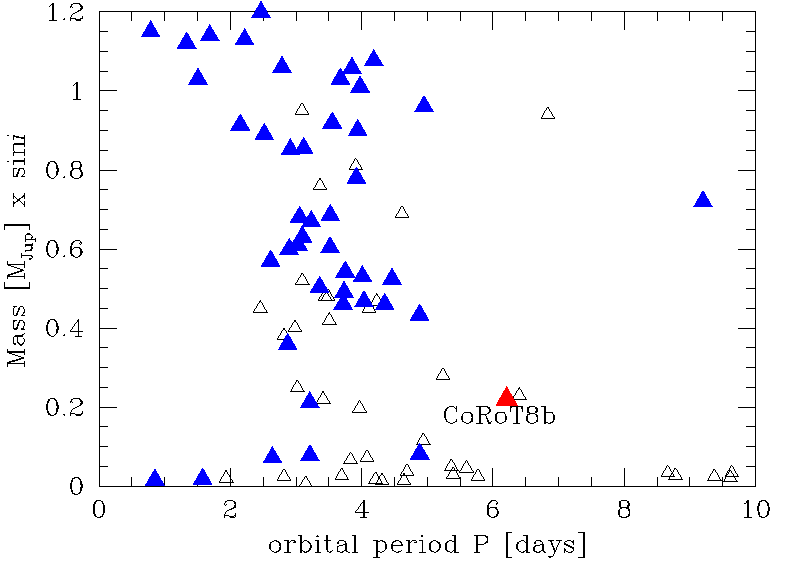}
      \caption{Correlation diagram between mass and period for close-in planets.
      CoRoT-8b appears to lie in between the bulk of giant planets and the bulk
      of super-Earths. Open triangles correspond to planets with unknown orbital
      inclinations.}
      \label{fig:mass-period}
   \end{figure}

   Radial velocity surveys do not find a lot of planetary companions at close distances ($P \le 10$~days)
   with masses comparable to CoRoT-8b. In the mass-period diagram (Fig.~\ref{fig:mass-period}),
   CoRoT-8b appears in between hot-Jupiter and super-Earth populations. This could challenge the existence
   of a bi-modal distribution \citep{Mordasini2009}, or CoRoT-8b might belong to the distribution tail of
   giant gaseous planets.
   
\begin{figure}
      \centering
      \includegraphics[width=8.5cm]{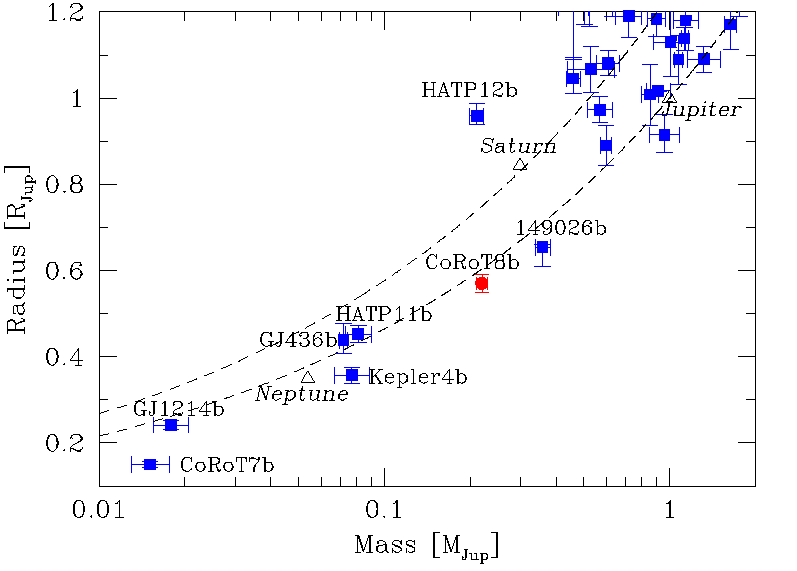}
      \caption{Mass-radius diagram for transiting planets including CoRoT-8b.
      The two dashed lines correspond to the densities of Saturn (0.69~\gcm3)
      and Jupiter (1.33~\gcm3). Neptune has a density of 1.76~\gcm3.}
      \label{fig:mass-radius}
   \end{figure}

   CoRoT-8b has a density of 1.6~\gcm3 (Fig.~\ref{fig:mass-radius}), which is higher than for
   Saturn (0.69~\gcm3), but comparable to Neptune (1.76~\gcm3) or HD\,149026\,b (1.7~\gcm3).
   The last orbits at 0.043~AU from its host star and has similar mass and radius to CoRoT-8b (0.36~\Mjup,
   0.65~\Rjup). By contrast, the density of HAT-P-12b (0.21~\Mjup, 0.96~\Rjup) that orbits at 0.038~AU
   from its host star is only 0.32~\gcm3.

   We estimate the thermal loss of hydrogen from CoRoT-8b by applying the method described by
   \cite{Lammer2009}. According to the expected evolution of the soft X-rays and EUV flux of K stars
   and CoRoT-8b's orbit location at 0.063~AU, the planet lost not more than about 0.045--0.12\,\%
   over an assumed integrated lifetime of 3~Ga. The lower mass-loss value corresponds to a heating
   efficiency of 10\,\% and the higher value to a heating efficiency of 25\,\%. From this estimation
   we can conclude that CoRoT-8b has only slightly evolved since its formation.

   To constrain the planetary composition, we combine stellar and planetary evolution
   models. The stellar evolution models are based on CESAM \citep{Morel2008}, include the
   diffusion of chemical elements in the star's radiative zone, and are calibrated to match our
   Sun for a mass of $1 \mathrm{M}_\odot$, a metallicity $[\mathrm{M}/\mathrm{H}]=0$, and an
   age of 4.5~Ga. A grid of models is computed as a function of mass, metallicity, and age. The
   two major constraints used are CoRoT-8's stellar density and effective temperature. For a
   given model, we define
   \begin{equation}
   n_{\sigma_\star} = \left[ \left(T_\mathrm{eff}(M_\star,[\mathrm{M}/\mathrm{H}],t_\star) \over
   \sigma_{T_\mathrm{eff}} \right)^2 + \left(\rho_\star(M_\star,[\mathrm{M}/\mathrm{H}],t_\star)
   \over \sigma_{\rho_\star} \right)^2 \right]^{1/2},
   \end{equation}
   which may be considered as the number of standard deviations from the combined
   $(T_\mathrm{eff},\rho_\star)$ constraints with corresponding standard deviations
   $(\sigma_{T_\mathrm{eff}}, \sigma_{\rho_\star})$. We look for models where $n_{\sigma_\star}$
   is small and at least less than 3. (The constraint on the spectroscopically-determined $\log g$
   value was found to be too weak to be useful.) Among these models, we select only the ones which
   have the right metallicity at the age considered, i.e., $0.2 \le [\mathrm{M}/\mathrm{X}](t) \le 0.4$.
   Since we account for chemical diffusion, the value of $[\mathrm{M}/\mathrm{X}](t)$ gets progressively
   lower with time. For a metal-rich star like CoRoT-8b, this favors slightly younger ages, but the star's
   low mass makes the effect very small.

   The corresponding constraints on the planetary parameters are then derived using the photometric
   and radial velocity constraints. Planetary evolution models are calculated with standard hypotheses  
   \citep[e.g.,][]{Guillot2008}: we assume that the planet is made of a central rock/ice core of
   variable mass and of an overlaying envelope of solar composition. Given that the planet is farther
   from its star than most other transiting planets, we do not consider the possibility of additional
   energy input from stellar tides.

\begin{figure}
      \centering
      \includegraphics[width=9cm]{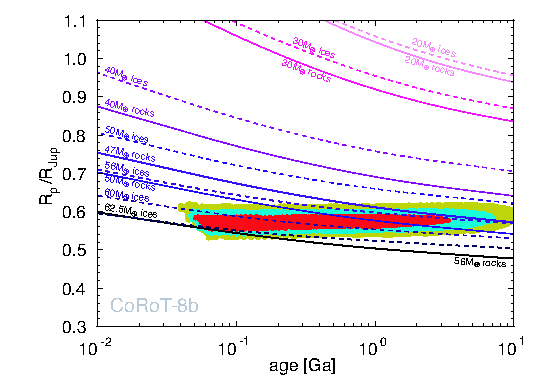}
      \caption{Evolution of the size of CoRoT-8b (in Jupiter units, 1~\Rjup = 71\,492~km) as a function
      of age (in Ga), compared to constraints inferred from CoRoT photometry, spectroscopy, radial
      velocimetry, and stellar evolution models. Red, blue, and green dots correspond to the planetary
      radii and ages that result from stellar evolution models matching the inferred
      $(\rho_\star,T_\mathrm{eff})$-uncertainty ellipse within $1 \sigma$, $2 \sigma$, and $3 \sigma$,
      respectively. Planetary evolution models for a planet with a solar-composition envelope over a
      central dense core of pure rocks and pure ices of variable mass are shown as plain and dashed lines,
      respectively. These models assume an equilibrium temperature of 925~K and a total mass of
      0.22~\Mjup. The inferred solutions at a $1 \sigma$ level are bounded by models with a minimum (rock)
      core mass of 47~$\mathrm{M}_\oplus$ and a maximum (ice) core mass of 62.5~$\mathrm{M}_\oplus$. The
      corresponding masses of the hydrogen-helium envelope are 22.9~$\mathrm{M}_\oplus$ (32.8\,\% M$\rm _{tot}$),
      and 7.4~$\mathrm{M}_\oplus$ (10.6\,\% M$\rm _{tot}$),
      respectively.}
      \label{fig:evolution}
   \end{figure}

   The results in terms of planetary size as a function of system age are shown in Fig.~\ref{fig:evolution}.
   The colored areas indicate the constraints derived from the stellar evolution models, for values
   of $n_{\sigma_\star}=1$, 2, and 3, respectively. The age constraints are relatively weak but favor
   values below 3 Ga. The planetary radius is tightly constrained, on the other hand. The amount
   of heavy elements that are found to bound the $n_{\sigma_\star}=1$ surface are 47 to 56~$\mathrm{M}_\oplus$ of
   rocks or 56 to 62.5~$\mathrm{M}_\oplus$ of ices.

   With a density comparable to that of Neptune, CoRoT-8b has a mass of heavy elements similar to that
   of HD~149026~b \citep{Sato2005,Ikoma2006}, but a much smaller hydrogen-helium envelope. As for  
   HD~149026, the parent star is metal-rich, confirming the trend that metal-rich stars tend to form
   giant planets with high contents in heavy elements \citep{Guillot2006,Burrows2007,Guillot2008}. The
   question of the formation of these planets with high masses in heavy elements but small hydrogen-helium
   envelopes remains open, but probably involves giant collisions \citep{Ikoma2006}.

\begin{acknowledgements}
   We wish to thank the French National Research Agency (ANR-08-JCJC-0102-01) for its continuous support
   for our planet-search program.
   The team at the IAC acknowledges support by grant ESP2007-65480-C02-02 of
   the Spanish Ministerio de Ciencia e Innovaci\'on.
   The German CoRoT Team (TLS and the University of Cologne)
   acknowledges DLR grants 50OW0204, 50O0603, and 50QP07011.
   T.~M. acknowledges the supported of the Israeli Science Foundation (grant no. 655/07).
   This research has made use of the SIMBAD database, operated at the CDS, Strasbourg, France,
   and of NASA's Astrophysics Data System.
\end{acknowledgements}


\bibliographystyle{aa}  
\bibliography{corot-8b} 

\end{document}